\documentclass[a4paper]{jpconf}
\usepackage{graphicx}
\usepackage{amsmath}
\usepackage{hyperref}
\usepackage{fancyhdr}

\def\to{\rightarrow}
\def\eps{\epsilon}

\begin{document}

CERN-PH-TH/2013-246

\title{Multi-loop Integrand Reduction with Computational Algebraic Geometry}

\author{Simon Badger$^1$, Hjalte Frellesvig$^{2,3}$, Yang Zhang$^3$}

\address{$^1$ Theory Division, Physics Department, CERN, CH-1211 Geneva 23, Switzerland}
\address{$^2$ Instituto Nazionale di Fisica Nucleare, Sezione di Roma,
P.le Aldo Moro 2, 00185 Roma, Italy}
\address{$^3$ Niels Bohr International Academy and Discovery Center,
The Niels Bohr Institute,\\%
University of Copenhagen, Blegdamsvej 17, DK-2100 Copenhagen, Denmark}

\ead{simon.badger@cern.ch}

\begin{abstract}
  We discuss recent progress in multi-loop integrand reduction
  methods. Motivated by the possibility of an automated construction of
  multi-loop amplitudes via generalized unitarity cuts we describe a
  procedure to obtain a general parameterisation of any multi-loop integrand in
  a renormalizable gauge theory. The method relies on computational algebraic
  geometry techniques such as Gr\"obner bases and primary decomposition of
  ideals. We present some results for two and three loop amplitudes
  obtained with the help of the \textsc{Macaulay2} computer algebra system and the
  Mathematica package \textsc{BasisDet}.
\end{abstract}

\section{Introduction}

There has been huge progress recently in the programme to fully automate
one-loop amplitude computations. Much of this success has been possible thanks to
the development of new reductions methods such as unitarity~\cite{Bern:1994zx,Bern:1994cg}, generalized unitarity~\cite{Britto:2004nc} and
integrand reduction~\cite{Ossola:2006us} which allow efficient generation of one-loop amplitudes
from simpler tree-level building blocks. We refer to G. Ossola's contribution to these proceedings
for more details of the status of this field \cite{Ossola:2013jea}.

There has been some interest in extending these techniques to multi-loop
amplitudes in the hope of improving our predictions of multi-particle
production beyond $2\to2$ scattering processes. Thanks to developments in
infra-red subtraction techniques full NNLO QCD computations with coloured initial and final
states have been now been achieved for $pp\to H+j$ \cite{Boughezal:2013uia}, $pp\to
t\bar{t}$ \cite{Czakon:2013goa} and $pp\to 2j$ \cite{Ridder:2013mf} which give hope that a similar level of precision
prediction may also be possible for $2\to3$ processes in the near future. In order to
combat the rapid growth of the Feynman diagrams in such computations, on-shell
techniques only work with physical degrees of freedom and simplify intermediate
steps.

The traditional approach to multi-loop Feynman diagram computations using
integration-by-parts identities has been well studied since its proposal more
than 30 years ago \cite{Chetyrkin:1981qh}. Several automated public codes have
been written and new developments are constantly increasing the range of
applicability, see for example the contribution of R. Lee to these proceedings
\cite{Lee:2012cn,Lee:2013mka}. Nevertheless complete solutions for multi-scale
problems are limited and are computationally intensive.

Since a complete integral basis for a general multi-loop amplitude is still
unknown it has been difficult to proceed with the conventional unitarity
construction of loop amplitudes. Nevertheless progress in understanding the
planar basis by explicit construction of unitarity compatible IBP relations
\cite{Gluza:2010ws,Schabinger:2011dz}.  This procedure has been a key
ingredient in the maximal unitarity method proposed by Kosower and Larsen
\cite{Kosower:2011ty}, where master integral coefficients are computed directly
from on-shell cuts into tree-level amplitudes after finding a specific contour of a
complex integral. The method has now been applied to a number of
high-multiplicity cases of the two-loop planar double
box~\cite{CaronHuot:2012ab,Johansson:2012zv,Johansson:2013sda} and also to
non-planar topologies~\cite{Sogaard:2013yga}.

The extension of the OPP integrand reduction method was first proposed by
Mastrolia and Ossola~\cite{Mastrolia:2011pr}. A key insight in how to construct
a general procedure has come from algebraic geometry. As we will see, the
problem reduces to operations on (quadratic) polynomial equations with many
variables. Such operations are well known in mathematics and many techniques,
and programs, now exist for dealing with computational algebraic geometry
problems. The most well known of these is Buchberger's algorithm for the
construction of Gr\"obner bases used in multivariate polynomial division. This
direction has been explored both in the context of reducing Feynman diagrams
\cite{Mastrolia:2012an,Mastrolia:2012wf,Mastrolia:2013kca} and using
generalized unitarity \cite{Badger:2012dp,Zhang:2012ce,Badger:2012dv}. More
formal applications of computational algebraic geometry have also bee used
to analyze the general structure of two and higher loop amplitudes \cite{Feng:2012bm,Huang:2013kh}.

In these proceedings we present a brief overview of the methods presented in refs.
\cite{Badger:2012dp,Zhang:2012ce,Badger:2012dv}. We describe the integrand reduction
procedure valid in $D$-dimensions for an arbitrary loop amplitude. We outline applications
at both two and three loops and consider the outlook for the future.

\section{Multi-loop integrand reduction with computational algebraic geometry}

The multi-loop amplitudes in this paper will refer to colour ordered primitive
amplitudes which have a fixed ordering for the external legs and a well defined
set of internal propagators. An $L$-loop amplitude in $D$ dimensions can be
schematically represented as,
\begin{align}
  A_n^{(L),[D]} = \int \prod_{i=1}^{L} \frac{d^{D} {k_i}}{(2\pi)^D}
  \frac{N(\{k\},\{p\})}{\prod_{l=1}^{L(L+9)/2} D_l(\{k\},\{p\})},
  \label{eq:integraldef}
\end{align}
where $D_l$ are the denominators appearing in the parent topology, $\{k\}$ are
the loop integration variables and $\{p\}$ are the external momenta. The
numerator function $N$ is a function of Lorentz products of loop momenta and
external momenta, external polarization vectors or external spinor
wave-functions. All kinematic quantities can be represented in terms of
invariants or scalar products of loop momenta with vectors in a basis
$\{e_1,\dots,e_4\}$ which can be constructed from the independent external
momenta plus additional spurious directions as proposed by van Neerven and Vermaseren
\cite{vanNeerven:1983vr}. We will write all scalar products as $k_i\cdot e_j =
x_{ij}$.  We must also include extra-dimensional variables $\mu_{ij} =
-k^{[-2\eps]}_i\cdot k^{[-2\eps]}_j$ where $k_i = \bar{k}_i + k_i^{[-2\eps]}$
and $\bar{k}_i$ is the four-dimensional components of $k_i$.
The number of propagators in $D=4-2\eps$ dimensions is $L(L+9)/2$ coming from
the $4L$ components of the $4$-dimensional loop momenta plus an additional
$L(L+1)/2$ components needed to describe the extra dimensional terms.

The integrand reduction method begins by reducing the numerator into terms proportional
to the denominators $D_l$ plus a remainder which is integrand of the topology in question,
\begin{equation}
  N(\{k\},\{p\}) = \Delta_{M}(\{k\},\{p\}) + \sum_{l=1}^{M} a_l D_l(\{k\},\{p\})
  \label{eq:ired1}
\end{equation}
where $M = \text{min}(n+3(L-1),L(L+9)/2)$ represents the maximum cut topology.
The sum of terms proportional to denominators $D_l$ will all vanish when
all propagators are put on-shell. To perform this operation we must first parametrise
the integrand $\Delta_M$ in terms of rational coefficients multiplying monomials of
scalar products,
\begin{align}
  \Delta_{M}(\{k\},\{p\}) = \sum_k c_k \times
  \prod_{i=1}^L\prod_{j=1}^{4} x_{ij}^{\alpha_{k;ij}}
                  \prod_{l=1}^L\prod_{m=l}^L (\mu_{lm})^{\beta_{k,lm}},
  \label{eq:ccoeffs}
\end{align}
where $c_i$ are rational functions of the external kinematics and the
$\alpha_{k;ij}$ are integer powers of the Lorentz products $x_{ij}$ and
$\beta_{k,lm}$ are the powers of the extra dimensional parameters $\mu_{lm}$.
The notation used in eq. \eqref{eq:ccoeffs} includes an over complete set of
scalar products (i.e. the matrices $\alpha_{k,ij}$ will contain many zeros).
For this reason it is necessary to distinguish reducible scalar products (RSPs),
which are linear in the propagators, from the irreducible scalar products (ISPs)
which satisfy non-linear identities. The basis of ISPs can be constructed from
the system of propagators as we will describe in the section \ref{sec:intparam}.

Having obtained this representation we can then proceed to compute the
coefficients $c_k$ from products of tree-level amplitudes. This requires solving the system
of on-shell constraints $D_l(\{k\},\{p\}) = 0$ which will in general have many
possible families of solutions $\{k^{(s)}\}$. Each of these solutions can be
parametrised in terms of a number of free variables,
$\{\tau_1,\ldots,\tau_{L(L+9)/2-M}\}$ \footnote{The existence of this parametrization is only guaranteed in $D=4-2\eps$ dimensions. In cases where the cut equations
have genus$>1$ this form will not exist (for example the 10-leg two-loop double box with 2 massless momenta at each of the 6 vertices).}. On each cut solution the integrand
factorises into a product of tree-level amplitudes that can be written as a
polynomial in $\tau_i$,
\begin{align}
  \Delta_{M}(\{k^{(s)}\},\{p\})
  &= \prod_{V=1}^{M-1} A^{(0)}(V,\{\{k^{(s)}\}\}),\\
  &= \sum_i d_{i} \prod_{j=1}^{L(L+9)/2-M} \tau_j^{\gamma_{k;j}},
  \label{eq:dcoeffs}
\end{align}
where we have used a schematic notation for the on-shell tree-level amplitude
appearing at each vertex of the cut-topology, $A^{(0)}(V,\{\{k^{(s)}\}\})$.
$\gamma_{k;j}$ are integer powers that follow from $\alpha_{k;ij}$ and
$\beta_{k;lm}$. Combining the information obtained in eqs. \eqref{eq:ccoeffs}
and \eqref{eq:dcoeffs} allows to construct a linear system relating the
rational coefficients of the integrand with the coefficients $d_i$ that can be
obtained from tree-level amplitudes,
\begin{equation}
  B\cdot \vec{c} = \vec{d},
  \label{eq:linearsystem}
\end{equation}
where $\vec{d} =
(d_1^{(1)},\ldots,d_{n(1)}^{(1)},d_1^{(2)},\ldots,d_{n(S)}^{(S)})$ and $n(s)$
is the dimension of the on-shell solution $s$ of which there are total of $S$
solutions. Solving this system then determines the coefficients $c_i$.

After the form of $\Delta_M$ has been determined we can proceed with lower
propagator cuts by performing the OPP style subtraction of the singularities,
\begin{align}
  \Delta_{m}(\{k^{(s)}\},\{p\}) =
    \prod_{V=1}^{m-1} A^{(0)}(V,\{\{k^{(s)}\}\})
    - \sum_{T=m+1}^{M}\sum_{t\in T} \frac{\Delta_{T;t}(\{k^{(s)}\},\{p\})}{\prod_{l \in P(t)/P(m)} D_l(\{k^{(s)}\},\{p\})}
\end{align}
where $T$ is the number of propagators in the higher topologies. For each multiplicity we must
sum over all the possible topologies $t$ each of which will have a set of additional propagators
in comparison with the topology $m$ in question. The sum is represented as $P(t)/P(m)$.

If we restrict to the case of $4$ dimensions $M=\text{min}(n+3(L-1),4L)$ and all
coefficients $\beta_{k,lm}=0$. There are two important details that need to be
clarified before we can show that such an approach will work:
\begin{enumerate}
  \item How do we define the integrand parameterisation? (i.e. compute $\alpha_{k,ij}$ and $\beta_{k,lm}$)
  \item How can we determine the number of on-shell solutions $S$ to define $\vec{d}$?
\end{enumerate}

\subsection{Integrand parameterisation \label{sec:intparam}}

The solution to first question above has been presented using multi-variate polynomial division with Gr\"obner
basis Zhang \cite{Zhang:2012ce} and also Mastrolia, Mirabella, Ossola and Peraro \cite{Mastrolia:2012an}.

In this section we will construct, using the language of algebraic geometry, an
ideal from the set of propagators for a topology with $m$ denominators,
\begin{equation}
  P = \{D_1,\cdots,D_m\}
  \label{eq:plist}
\end{equation}
whose integrand we will denote as $\Delta_m$. Using the van Neerven Vermaseren
construction we are able to construct a basis where external momenta
$\{p_1,\ldots,p_x\}$ span the physical space and the directions
$\{\omega_1,\cdots,\omega_4\}$, orthogonal to the physical space, span the
spurious, or trivial, space:
\begin{equation}
  e = \{e_1,\cdots,e_4\}.
\end{equation}
All of the Lorentz products appearing in the expansion of $P$ can be written in terms
of the basis products $k_i\cdot e_j = x_{ij}$ using the $4\times4$ Gram matrix $[G_4]_{ij} = e_i\cdot e_j$,
\begin{equation}
  k_i \cdot v_j = (k_i\cdot e_1\: k_i\cdot e_2\: k_i\cdot e_3\: k_i\cdot e_4).
  G_4^{-1}.
  \begin{pmatrix}
    v_j\cdot e_1 \\
    v_j\cdot e_2 \\
    v_j\cdot e_3 \\
    v_j\cdot e_4
  \end{pmatrix}
  \label{eq:gramid}
\end{equation}
and
\begin{equation}
  k_i \cdot k_j = -\mu_{ij} + (k_i\cdot e_1\: k_i\cdot e_2\: k_i\cdot e_3\: k_i\cdot e_4).
  G_4^{-1}.
  \begin{pmatrix}
    k_j\cdot e_1 \\
    k_j\cdot e_2 \\
    k_j\cdot e_3 \\
    k_j\cdot e_4
  \end{pmatrix}
  \label{eq:gramid2}
\end{equation}
The above assumes that we have all external vectors in four dimensions and hence only obtain contributions from
the $4D$ parts of the loop momenta $k_i\cdot e_j = \bar{k}_i\cdot e_j$. To find the constraints from the propagators
on the integrand we will use eqs. \eqref{eq:gramid} and \eqref{eq:gramid2} to translate eq. \eqref{eq:plist} into
polynomials in $x_{ij}$. The RSPs could be identified from linear relations among the equations however an efficient
method is to compute the Gr\"obner basis using a graded monomial ordering \cite{Zhang:2012ce}.

\begin{figure}[h]
  \begin{center}
    \includegraphics{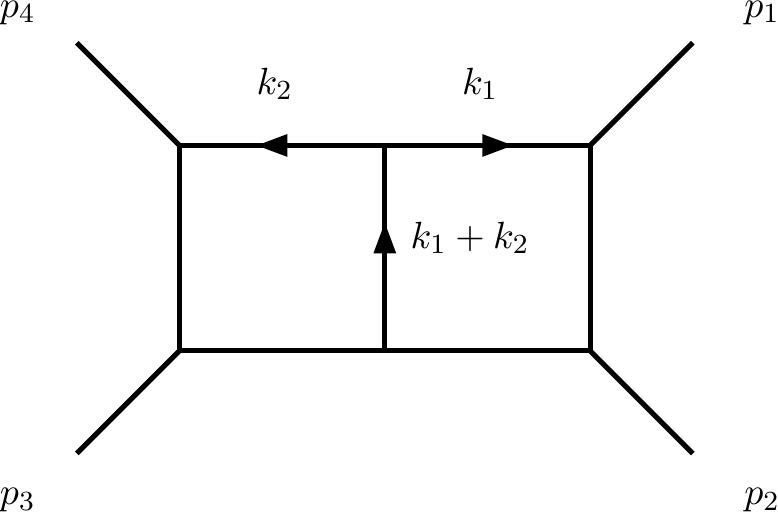}
  \end{center}
  \caption{The two-loop planar double box configuration}
  \label{fig:doublebox}
\end{figure}

\begin{figure}[h]
  \begin{center}
    \includegraphics{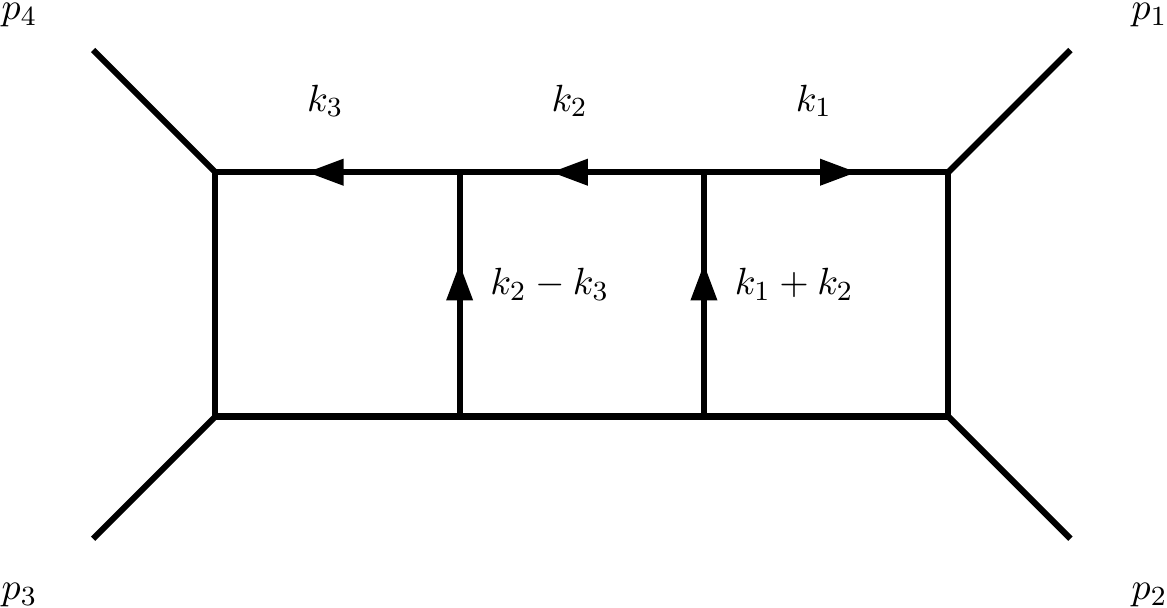}
  \end{center}
  \caption{The three-loop triple box configuration}
  \label{fig:triplebox}
\end{figure}

It is helpful to consider an example at this point so we take the case of the two-loop planar double box in four dimensions.
The list of propagators is,
\begin{equation}
  P = \{k_1^2, (k_1-p_1)^2, (k_1-p_1-p_2)^2, k_2^2, (k_2-p_4)^2, (k_2-p_3-p_4)^2, (k_1+k_2)^2\}
  \label{eq:2ldbprops}
\end{equation}
and we will span $k_1$ and $k_2$ with the basis,
\begin{equation}
  \vec{e} = (p_1, p_2, p_4, \omega)
  \label{eq:2ldbbasis}
\end{equation}
where $\omega^\nu = 4i\varepsilon^{\nu 1 2 4}$. The Gram matrix is explicitly
\begin{equation}
  G_4 = \frac{1}{2}
  \begin{pmatrix}
    0 & s & t & 0 \\
    s & 0 & u & 0 \\
    0 & u & t & 0 \\
    0 & 0 & 0 & stu \\
  \end{pmatrix}
  \label{eq:dbgram}
\end{equation}
where $u = -s-t$. It is then straightforward to apply eq. \eqref{eq:gramid} and express $P$ in terms
of the 8 scalar products $x_{ij}$. We find four linear relations in the on-shell equations:
\begin{align}
  x_{11} &= 0 & 2 x_{12} + s &= 0 &
  x_{23} &= 0 & 2 x_{22} + 2 x_{21} + s &= 0
  \label{eq:dbRSPeqs}
\end{align}
which leaves us with a set of $4$ ISPs $\{x_{24}, x_{14}, x_{21}, x_{13}\}$ and an ideal
\begin{align}
  I = \big\langle &-4 t x_{13}+4 x_{13}^2-4 x_{14}^2+t^2,-4 t x_{21}+4 x_{21}^2-4 x_{24}^2+t^2,\nonumber\\&
  s t \left(2 x_{13}+2 x_{21}\right)+s \left(4 x_{13} x_{21}-4 x_{14} x_{24}\right)+8 t x_{13} x_{21}-s t^2 \big\rangle
  \label{eq:dbideal}
\end{align}
which we should use to remove terms from the polynomial ansatz for $\Delta_7$ restricted by imposing
renormalizability constraints,
\begin{equation}
  \Delta_7^{\text{ansatz}} = \sum_{a_1,\dots,a_4} \mathcal{C}_{a_1 a_2 a_3 a_4} \, x_{14}^{a_1} x_{24}^{a_2} x_{13}^{a_3} x_{24}^{a_4}
\end{equation}
where $a_1+a_3 \leq 4, a_2+a_4 \leq 4$ and $a_1 + a_2 + a_3 + a_4 \leq 6$ which is a sum of 160 terms.
The polynomial division operation based on Gr\"obner bases is a standard algorithm in computer algebra
packages and so it is now straightforward to obtain:
\begin{align}
  \Delta_7^{\text{ansatz}} / G(I) = \Delta_7 = \sum_{i=1} c_i \, x_{14}^{\alpha_{i1}} x_{24}^{\alpha_{i2}} x_{13}^{\alpha_{i3}} x_{21}^{\alpha_{i4}}
\end{align}
which contains 32 terms defined by:
\begin{align}
  \alpha = (&(0, 0, 0, 0), (0, 0, 0, 1), (0, 0, 0, 2), (0, 0, 0, 3), (0, 0, 0, 4), \nonumber\\&
(0, 0, 1, 0), (0, 0, 1, 1), (0, 0, 1, 2), (0, 0, 1, 3), (0, 0, 1, 4), \nonumber\\&
(0, 0, 2, 0), (0, 0, 2, 1), (0, 0, 3, 0), (0, 0, 3, 1), (0, 0, 4, 0), \nonumber\\&
(0, 0, 4, 1), (0, 1, 0, 0), (0, 1, 0, 1), (0, 1, 0, 2), (0, 1, 0, 3), \nonumber\\&
(0, 1, 1, 0), (0, 1, 1, 1), (0, 1, 1, 2), (0, 1, 1, 3), (0, 1, 2, 0), \nonumber\\&
(0, 1, 3, 0), (0, 1, 4, 0), (1, 0, 0, 0), (1, 0, 0, 1), (1, 0, 1, 0), \nonumber\\&
(1, 0, 2, 0), (1, 0, 3, 0))
\end{align}
The algorithm presented here is implemented in the public Mathematica package
{\sc BasisDet} by Yang Zhang. We note the the problem of determining the integrand
parameterisation is rather efficient, especially when external kinematic
quantities are computed numerically, and has been applied to many complicated
multi-loop topologies in $4$ and $4-2\eps$ dimensions. The integrand basis can
be rather large for higher loop topologies. The three-loop triple box, shown in figure \ref{fig:triplebox},
considered in ref.~\cite{Badger:2012dv} contains a total 398 monomials in terms
of 7 ISPs. In this case IBP reduction was used subsequently to reduce the
integrand onto a basis of three master integrals.

\subsection{On-shell solutions \label{sec:onshell}}

The solution to the second question has been formulated by Zhang in ref.
\cite{Zhang:2012ce}. In general the geometry of the solutions space can be very
complicated but algebraic geometry again provides a technique to systematically
analyze the zeros of the propagator ideal $I$, which will denote
$\mathcal{Z}(I)$ or the zero locus of $I$.  The famous Lasker-Noether theorem for
primary decomposition of ideals proves that any ideal may be decomposed into a
set of smaller primary ideals,
\begin{equation}
  I = \bigcap\limits_{s=1}^{S} I_s
  \label{eq:pdecomp}
\end{equation}
where ${\rm dim}(I_s) = n(s)$. The zero locus of $I$ follows from this since,
\begin{equation}
  \mathcal{Z}(I) = \bigcup\limits_{s=1}^{S} \mathcal{Z}(I_s).
  \label{eq:Zpdecomp}
\end{equation}
Algorithms to compute the primary decomposition are available in the
\textsc{Macaulay2} computer algebra package \cite{M2}.  For the two loop double box
considered in the previous section one can use this technique to reproduce the
6 independent solutions first obtained by Kosower and Larsen
\cite{Kosower:2011ty}. The three-loop triple box has a total of 14 solutions
and explicit parameterisation have been presented in ref.~\cite{Badger:2012dv}.

\subsection{$D$-dimensional cuts \label{sec:ddcuts}}

One more aspect of the linear system in eq. \eqref{eq:linearsystem} that
remains is to prove that the matrix $B$ does indeed have the property ${\rm
rank}(M) = {\rm dim}(\vec{c})$. This is not guaranteed in four dimensions.
However, as shown recently in ref. \cite{Badger:2013gxa}, by going to $D$
dimensions this problem is avoided at the cost of creating larger linear systems
relating the integrand coefficients to the tree-level amplitudes. We have
observed two cases where the procedure can break down:
\begin{enumerate}
  \item The propagator ideal is not radical, often written $I \neq \sqrt{I}$. In this case
    there will be terms in $\Delta$ that cannot be determined from the product
    of trees as they will vanish on each cut solution $\{k^{(s)}\}$.
  \item The dimension of a branch of the cut solutions is not regular. For example, in four dimensions,
    we can find that $\dim(I_i) \neq 4L - m$ for all primary ideals in an $m$ propagator cut. The two-loop
    pentagon-triangle is an example of this: though there are seven propagators we find two families of
    solutions with dimension 2 instead of dimension 1.
\end{enumerate}
There is a possibility these problems could be avoided by considering different
topologies simultaneously but it will complicate the integrand reduction
procedure. In $4-2\eps$ dimensions however this problem is side stepped since
one can prove that all ideals are radical ideals \cite{Badger:2013gxa}. In this
case there will always be exactly one solution to the on-shell constraints
($S=1$).

In order to generate the extra-dimensional dependence in the tree amplitudes it
is necessary at two-loops to use at least six dimensions. In general tree amplitudes
will need to be computed in higher and higher dimensions as we increase the loop
order.

\section{Outlook}

We have presented an overview of an integrand reduction procedure valid in $D$
dimensions which allows the rational coefficients of the integrand monomials to
be extracted from tree-level amplitudes. The method has been applied in the
recent computation of the planar all-plus helicity five-gluon two-loop
amplitude, the first example of a $2\to3$ helicity amplitude in QCD \cite{Badger:2013gxa}.

There are still many questions as to whether the method can be an efficient
alternative to integration-by-parts identities in the future. Of course the
integrand basis can always be further reduced by IBP to reduce the number of
integrals that must be computed. There is also an open question concerning the
treatment of doubled propagator diagrams that can appear in the input. In this
case it has been shown that there is no obstacle in performing the polynomial
division \cite{Mastrolia:2013kca} yet the on-shell cuts into tree amplitudes
will need regulating in order to follow the approach suggested here.

The use of algebraic geometry in multi-loop computations may have many
applications in the future both for integrand computations and in the
computations of the integrals themselves. Progress in the computation of the
loop integrals themselves, for example the differential equation of Henn
\cite{Henn:2013pwa}, give hope that precision predictions of multi-particle
states at NNLO may be possible within the near future.

\section*{References}


\providecommand{\newblock}{}

\end{document}